\documentclass[conference]{IEEEtran}
\IEEEoverridecommandlockouts

\usepackage{cite}
\usepackage{amsmath,amssymb,amsfonts}
\usepackage{algorithm, algorithmic}             
\usepackage{graphicx}
\usepackage{textcomp}
\usepackage{xcolor}
\usepackage{mathtools}
\usepackage{mathrsfs}  
\usepackage{multirow}
\usepackage{multicol}
\usepackage{makecell}
\usepackage{subfig}
\usepackage{url}

\usepackage{amsthm}
\theoremstyle{definition}

\def\BibTeX{{\rm B\kern-.05em{\sc i\kern-.025em b}\kern-.08em
    T\kern-.1667em\lower.7ex\hbox{E}\kern-.125emX}}

\begin{document}

\title{DSO: A GPU Energy Efficiency Optimizer by Fusing Dynamic and Static Information}

\author{
\IEEEauthorblockN{Qiang Wang\IEEEauthorrefmark{1}, Laiyi Li\IEEEauthorrefmark{1}, Weile Luo\IEEEauthorrefmark{2}, Yijia Zhang\IEEEauthorrefmark{3}, Bingqiang Wang\IEEEauthorrefmark{3}\thanks{Corresponding authors: Qiang Wang, Yijia Zhang}}

\IEEEauthorblockA{\IEEEauthorrefmark{1}Harbin Institute of Technology (Shenzhen)}
\IEEEauthorblockA{\IEEEauthorrefmark{2}The Hong Kong University of Science and Technology (Guangzhou)}
\IEEEauthorblockA{\IEEEauthorrefmark{3}Peng Cheng Laboratory}

\IEEEauthorblockA{\IEEEauthorrefmark{1}qiang.wang@hit.edu.cn, \IEEEauthorrefmark{1}23s151122@stu.hit.edu.cn, \IEEEauthorrefmark{2}wluo976@connect.hkust-gz.edu.cn, \IEEEauthorrefmark{3}\{zhangyj01,wangbq\}@pcl.ac.cn}

}

\maketitle
\begin{abstract}
  Increased reliance on graphics processing units (GPUs) for high-intensity computing tasks raises challenges regarding energy consumption. To address this issue, dynamic voltage and frequency scaling (DVFS) has emerged as a promising technique for conserving energy while maintaining the quality of service (QoS) of GPU applications. However, existing solutions using DVFS are hindered by inefficiency or inaccuracy as they depend either on dynamic or static information respectively, which prevents them from being adopted to practical power management schemes. To this end, we propose a novel energy efficiency optimizer, called DSO, to explore a light weight solution that leverages both dynamic and static information to model and optimize the GPU energy efficiency. DSO firstly proposes a novel theoretical energy efficiency model which reflects the DVFS roofline phenomenon and considers the tradeoff between performance and energy. Then it applies machine learning techniques to predict the parameters of the above model with both GPU kernel runtime metrics and static code features. Experiments on modern DVFS-enabled GPUs indicate that DSO can enhance energy efficiency by 19\% whilst maintaining performance within a 5\% loss margin.
\end{abstract}




\maketitle

\section{Introduction}

In recent years, there has been a significant increase in the utilization of graphics processing units (GPUs) as accelerators in high-performance computing (HPC). This trend has been driven by the growing demand for energy-efficient solutions, particularly due to the rise of artificial intelligence (AI) applications \cite{zeus_nsdi2023,tpds2022_scheduling,dlw_sc2021}. One prominent instance is the advanced language model GPT-3 \cite{floridi2020gpt}, which was designed with over 150 billion parameters to generate human-like texts. The training cost for GPT-3 exceeds 4.6 million dollars, equivalent to nearly 120 years of electricity consumption by an average household.
These staggering figures underscore the significance of implementing effective mechanisms to enhance the energy efficiency of these systems. Even a modest 5\% reduction in energy consumption can have a substantial impact.

Dynamic voltage and frequency scaling (DVFS) is a promising technique for GPUs which enables the adjustment of devices to lower performance/power states. DVFS optimizes GPU performance and power by adjusting voltage and frequency levels, offering substantial energy savings with minimal performance impact \cite{powerS2016,survey2017}. Recent investigations \cite{tang2019dvfs_dl,dlw_sc2021} demonstrate that the utilization of DVFS techniques in graphics processing units (GPUs) engaged in deep neural network (DNN) applications resulted in energy savings of up to 26\% especially in the common DNN inference scenario.
Several existing studies \cite{amd_hpca2015,wang2018_icpads,dvfs_aware_parco2019,nn_access2019,tpds2020_gpudvfs} on energy conservation through DVFS rely on runtime information provided by GPU profiling tools, such as \textsl{nvprof} for Nvidia GPUs. These tools have proven to be effective in modeling the performance and power changes under different DVFS settings, given the high correlation between the performance counters of each GPU sub-component and the execution time/power. However, two drawbacks hinder their practical online usage. 
Firstly, the profiling overhead associated with these tools is typically significant because these profiling tools often require multiple replays of the target application, resulting in heavy computational costs.
Secondly, some of these tools necessitate modifications to the application source code, which is not user-friendly and may not be available for online submitted jobs. 

Another branch of analyzing the performance and power behavior of GPU applications is static information modeling, which involves examining GPU low-level assembly codes such as PTX\footnote{\url{https://docs.nvidia.com/cuda/parallel-thread-execution/index.html}} and SASS\footnote{\url{https://docs.nvidia.com/cuda/cuda-binary-utilities/}}. This approach relies on using the GPU assembly of the kernels, which can be obtained at compile-time or by disassembler tools. One advantage of this approach is that it does not require modifying users' applications or pre-executing them to collect runtime information. Moreover, this type of static modeling introduces new usage scenarios, such as facilitating the evaluation of how changes in the source code can affect the DVFS behavior of applications. However, due to the lack of GPU runtime information such as cache hit rate and compute resource occupancy, the prediction errors for execution time are typically high.

We argue that an ideal energy efficiency optimizer for DVFS-based GPUs should be efficient and accurate, which cannot tolerate the extremely high overhead of those existing profiling tools. Recently, data center GPU manager (DCGM) \cite{dcgm} published by Nvidia is a lightweight tool to manage and monitor the GPUs in data center environments. It provides a set of powerful tools and APIs that allow system administrators and operators to monitor the health, performance, and utilization of the GPUs in a non-intrusive manner with negligible cost. However, the metrics provided by DCGM is just a subset of those by \textsl{nvprof}, which may decrease the model accuracy. 
To this end, we finally come to the optimization framework, called DSO, that leverages both the DCGM metrics (dynamic information) and the PTX codes (static information). We summarize the contributions of DSO as follows.
\begin{itemize}
    \item We propose a novel parameterized theoretical model of GPU energy efficiency considering both the effects of DVFS and the tradeoff between performance and energy consumption. The optimization solution for a GPU kernel is also explicitly derived to tune the best DVFS setting.
    \item We design a machine learning based scheme to predict the parameters of the proposed theoretical model leveraging both the runtime metrics from the lightweight DCGM profiling tool and the static features from the PTX codes.
    \item Validated on 20 real applications (not used during training) among two contemporary GPUs, the model trained on only micro-benchmarks shows considerably low errors (mostly within 5\%) for both performance and power prediction. The average energy conservation observed in our optimization results achieves 19\% on Tesla V100 on average compared to the default setting with no more than 5\% performance loss, all without heavy offline profiling.
\end{itemize}
\section{Background and Motivation} \label{sec:BM}
\subsection{GPU DVFS}

Performance and power modeling is essential for energy conservation in different DVFS settings, dictating the total energy consumption of GPU applications. Recent studies \cite{powerS2016,survey2017,tang2019dvfs_dl} have shown that GPU DVFS behaviors are more complex than CPUs when altering voltages and frequencies, sometimes even proving contrary to conventional CPU DVFS. As modern GPUs have two primary frequency domains—core frequency for stream multiprocessors (SMs) speed and memory frequency for GPU memory bandwidth, our efforts centralize on optimizing frequency control based on application behaviors.

\subsection{The Input Sources for GPU Modeling}
Two input types are generally used in previous works for GPU performance, power modeling, and DVFS management: dynamic and static information.
Dynamic information, also referred to runtime information, is hardware-dependent, collected during one execution of the target application, usually through profiling tools like \textsl{nvprof} \cite{NVProfiler}. Despite high prediction accuracy, the profiling overhead may inhibit practical DVFS energy optimization due to the lack of real-time measurements.
Static information is hardware-independent, referring to GPU code features obtained before kernel execution. The NVIDIA Parallel Thread Execution (PTX), an intermediate assembly language for NVIDIA GPUs, is often utilized. It can be extracted from CUDA binary files using the Nvidia disassembler tool \textsl{cuobjdump}, allowing analysts to link each instruction with the GPU components involved in its execution.
Moreover, lightweight monitoring tools like \textsl{nvidia-smi} \cite{nvidia-smi} and DCGM \cite{dcgm} are available for tracing GPU status. As they relate to GPU hardware runtime, they fall under the category of runtime information.
\subsection{Why Dynamic and Static Information Fusion}
As stated above, these two input source types have their own advantages and drawbacks. A practical energy management scheme with GPU DVFS should be efficient and accurate in terms of modeling and optimization. Notice that DCGM can monitor the runtime utilization of different GPU components with negligible overhead. We consider it as the substitute of the heavy profiling tool such as \textsl{nvprof}. To fulfill the information gap between DCGM and \textsl{nvprof}, we further utilize the PTX code features as complement when designing our energy efficiency optimizer, which finally comes to the scheme of dynamic and static information fusion.
\section{Related Work} \label{sec:RW}

There have been many research studies about understanding the impacts on GPU performance and energy consumption brought by DVFS as well as optimizing them solely or jointly. They generally take advantage of either physical runtime metrics (also referred as dynamic information) and GPU kernel codes (also referred as static information).

Dynamic information-based methods initially used micro-benchmarks to profile GPU hardware components \cite{gpumembench2016,mixbench2017,mei2017tpds}. DVFS-aware prediction models were developed by Wang et al. \cite{tpds2020_gpudvfs} to quantify each GPU component's performance contribution. Guerreiro et al. employed similar tactics for power contributions \cite{guerreiro_hpca2018,guerreiro_tpds2019}. ML-based methods, such as one developed by Wang et al. \cite{gpgpu2020_cross}, encapsulated interplay influences of diverse instructions and achieved below 10\% average errors. The high accuracy of the above methods indicates the importance of runtime information, especially for contemporary GPU architectures with rich sub-components.

Static information-based methods began with Hong et al. \cite{mwp_isca2009,isca2010_hong}, who modelled GPU kernel performance and power consumption using static CUDA and PTX code analysis. Guerreiro et al. \cite{ptx2019} expanded this by including GPU assembly instruction sequences and deploying recurrent neural networks to capture dependency features. Braun et al. \cite{taco2021_portable} proposed a simple model for swift predictions across GPUs relying solely on PTX instruction statistics. Fan et al. \cite{icpp2019_predictable} developed DVFS-aware static models based on a vector of 10 instruction types. However, these methods achieved high power prediction accuracy, but their execution time predictions were lacking due to the absence of hardware runtime information, such as cache hit rate or register spilling. 

\section{Methodology} \label{sec:ml_framework}

\subsection{Overview}
\begin{figure}[ht]
    \centering
    \includegraphics[width=0.4\textwidth]{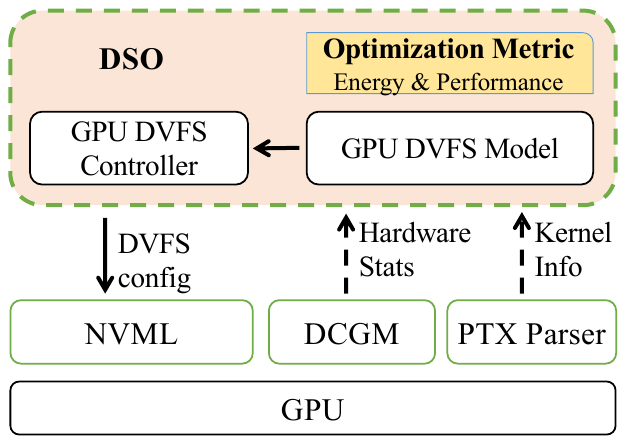}
    \caption{DSO Workflow.}
    \label{fig:overview}
\end{figure}

Figure \ref{fig:overview} outlines our DSO optimizer framework. Initially, we present a parameterized optimization model, referred to as the ``GPU DVFS Model" rectangle, which incorporates the effects of GPU DVFS and balances the tradeoff between performance and energy efficiency. Our model aims to enhance energy efficiency by considering these factors.
To accurately and efficiently predict the model parameters, we employ machine learning techniques that leverage hardware status information from DCGM and GPU kernel details from the PTX parser. By utilizing these inputs, we can make informed predictions about the model parameters. Once these parameters are determined, we can theoretically derive the optimal DVFS configuration.
For implementing the DVFS configuration, the GPU DVFS controller utilizes the APIs provided by NVML \cite{NVML}. These APIs enable us to set the desired voltage and frequency targets, allowing us to implement the optimal DVFS configuration based on the calculated parameters.

\subsection{Problem Formulation}\label{subsec:formulation}
Previous studies have showcased the effectiveness of statistical models trained using dynamic and static GPU features to accurately represent performance and power data samples. These models have achieved a remarkable level of accuracy and confidence. However, compared to learning-based approaches, the utilization of parameterized models provides the advantage of interpreting the unique characteristics of GPU hardware and comprehending the impact of DVFS on performance and power.
Expanding upon the research presented in \cite{tpds2022_scheduling,micro_2000}, we adopt a similar approach to model the runtime power of the GPU, as shown in Equation \eqref{eq:power_model}. 
\begin{align}
P(V^{c}, f^{c}, f^{m}) = \underbrace{(P^0+\kappa V^{c})}_{P_{static}} +
	\underbrace{(\gamma f^{m} + c(V^{c})^2 f^{c})}_{P_{dynamic}} \label{eq:power_model}
\end{align}

$V^{c}$, $f^{c}$, and $f^{m}$ represent the GPU supply core voltage, GPU core frequency, and GPU memory frequency, respectively. The power is consist of two parts, the static part $P_{static}$ and the dynamic part $P_{dynamic}$. $P_{static}$ includes $P^{0}$, which denotes the constant power consumption of the GPU system that is not related to GPU voltage/frequency scaling, and $\kappa V$, which denotes the power that maintains the supply voltage for executing the GPU application. $\kappa$ denotes the coefficient related to the hardware characteristics, such as the number of transistors in the chip design and the leakage current for a single transistor. As for the dynamic part, the coefficients $\gamma$ and $c$ are constant values that rely on both the hardware characteristics and the specific application being considered. These coefficients indicate the sensitivity of power consumption to memory frequency scaling and core voltage/frequency scaling, as explained in \cite{isca2010_hong}.

Compared to power modeling, performance modeling of GPU DVFS is rather complex \cite{nath2015crisp,tpds2020_gpudvfs}. Inspired by the DVFS-aware roofline observations in \cite{tpds2020_gpudvfs}, we innovatively design a piecewise mathematical model with a concise form to simplify the subsequent analysis of GPU energy conservation. We formulate the performance function $T(f^{c},f^{m})$ of a GPU-accelerated application as shown in Eq. \eqref{eq:time_model}. $t^0$ represents the constant component in GPU application execution time. $\alpha$ is a constant factor that indicates the sensitivity of this application to GPU memory frequency scaling, and $\beta$ is a constant factor that indicates the sensitivity to GPU core frequency scaling. With $t^0,\alpha$ and $\beta$ set to different values, the model is capable of simulating the various DVFS effects of a variety of applications. Our experiments on real GPU applications indicate that this time model effectively captures the performance effects of DVFS on all the tested applications. It provides a coherent explanation for the observed performance variations and accurately represents the impact of DVFS across different application scenarios.
\begin{align}
    T(f^{c},f^{m})=t^0+\text{max}(\frac{\alpha}{f^m}, \frac{\beta}{f^{c}})
    \label{eq:time_model}
\end{align}

Notice that $f^{c}$ and $V^{c}$ are correlated. For a fixed $V^{c}$, the maximum core frequency ($f^{c}_{max}$) is determined by $V^{c}$. We apply the function in \cite{hotpower_2013} to denote this relationship: $f^{c} \leq g_1(V^{c})=\sqrt{(V^c-\kappa)/2}+\kappa$, and $\kappa$ is the function parameter.

With the above models,
the GPU energy ($E_J$) consumed to process one task is the product of the runtime power and the execution time, as shown in Eq. \eqref{eq:energy_model}. 
\begin{align}
	E = \underbrace{(P^0+\kappa V^{c}+\gamma f^{m} + c(V^{c})^2 f^{c})}_{P(V^{c}, f^{c}, f^{m})} \times
	\underbrace{(t^0+\text{max}(\frac{\alpha}{f^m}, \frac{\beta}{f^{c}}))}_{T(f^{c})} \label{eq:energy_model}
\end{align}

We propose a simple objective cost function to tradeoff the performance and energy consumption as Eq. \eqref{eq:cost_function}.
\begin{align}
	C(V^{c}, f^{c}, f^{m}) &= \eta E + (1 - \eta) P_{max} T \nonumber \\
	&= \eta PT + (1-\eta)P_{max}T \nonumber \\
	&= (\eta P + (1-\eta)P_{max})T \label{eq:cost_function}
\end{align}

Here $\eta$ is the parameter specified by the user to express the relative importance of energy efficiency and training performance (throughput). When $\eta=0$, we are only optimizing for time consumption, whereas when $\eta=1$, we are only optimizing for energy consumption. $P_{max}$ is the maximum power limit supported by the GPU, a constant introduced to unify the units of measure in the cost metric.

The parameters to be determined are $(P^0, \kappa, \gamma, c)$ related to power and $(t^0, \alpha, \beta)$ related to performance. While they can be fitted with the data points sampled from different DVFS settings, in the paper, we attempt to directly predict them by the DCGM metrics and static code information, which do not need to pre-execute the target GPU application with different DVFS settings.

\subsection{Best Configuration for Target Cost}\label{subsec:solution}
As a first step, we consider the following how to derive the solution to achieve the best energy efficiency modeled by Eq. \eqref{eq:cost_function}. Eq. \eqref{eq:argmin_e} shows the mathematical formulation of the problem. Notice that $V^c$ and $f^c$ are correlated variables, and $f^c$ is upper bounded by a function of $V^c$, denoted by $g_1(V_c)$.

\begin{align} \label{eq:argmin_e}
    \text{argmin }C =&\text{argmin}\{(\eta(P^{0}+\kappa V^c+\gamma f^{m}+c(V^{c})^2f^{c}) \nonumber \\
    &+(1-\eta)P_{max}) \times (t^0+\text{max}(\frac{\alpha}{f^m}, \frac{\beta}{f^{c}}))\} \nonumber \\
    s.t. V^{c}_{min} &\leq V^{c} \leq V^{c}_{max}, f^{m}_{min} \leq f^{m} \leq f^{m}_{max},\nonumber \\
    f^{c}_{min} &\leq f^{c} \leq g(V^{c})
\end{align}

\noindent \textbf{Theorem 1.} With a fixed memory frequency, the cost function of a GPU kernel is minimum when the GPU core frequency is maximum corresponding to the GPU core voltage, and $f^c \leq \frac{\beta}{\alpha}f^{m}$, i.e.,

\begin{align}
    C_{min}(f^{m})=\underset{V^{Gc}}{\text{arg min}}C(V^{c},g(V^{c}),f^{m}). \nonumber
\end{align}

\begin{proof}
We firstly discuss the case of $f^c < \frac{\beta}{\alpha}f^{m}$ (equivalent to $\frac{\alpha}{f^m} < \frac{\beta}{f^{c}}$), which indicates that the kernel is compute-bound. The cost function then becomes $C=(\eta(P^0+\kappa V^{c}+\gamma f^{m}+c(V^{c})^2f^{c})+(1-\eta)P_{max}) \times (t^0+\frac{\beta}{f^c})$. We obtain the first-order partial derivatives as: $\frac{\partial C}{\partial V^{c}}=\eta(\kappa+2c V^{c} f^{c})(t^0+\frac{\beta}{f^c})$ and $\frac{\partial C}{\partial f^{c}}=c\eta (V^{c})^2(t^0+\frac{\beta}{f^c})-\beta\frac{\eta(P^0+\kappa V^{c}+\gamma f^{m}+c(V^{c})^2f^{c})+(1-\eta)P_{max}}{(f^{c})^2}$. Because $\frac{\partial C}{\partial V^{c}}>0$, $C$ cannot attain its minimum on the interior of the domain, and $E_J$ is a monotonically increasing function of $V^{c}$. The minimum is on the boundary of $g(V^{c})$. 

We then discuss the case of $f^c \ge \frac{\beta}{\alpha}f^{m}$, which indicates that the kernel is memory-bound. $T(f^c)$ is reduced to be $t^0+\frac{\beta}{f^m}$. As $P$ is a monotonically increasing function of $f^c$, the minimum of $C$ is achieved when $f^c = \frac{\beta}{\alpha}f^{m}$. 

To be concluded, $f^{c}$ can be eliminated such that finding the minimum of $C$ is only related to $V^{c}$, and the condition of getting the minimum $C$ is $f^c \leq \frac{\beta}{\alpha}f^{m}$.
\end{proof}

Theorem 1 transforms a three-variable optimization problem into a two-variable optimization problem. It implies that when we scale the GPU core alone to conserve energy, we only need to find an appropriate core voltage and set the core frequency to the largest allowed value.

We then consider GPU memory frequency scaling alone. If the core voltage and frequency settings are fixed as $V^{c}_o$ and $f^{c}_o$, we can easily compute the optimal memory frequency by setting $\frac{\partial C}{\partial f^c}=0$. We denote it as $f^{m}_o$. Since the minimum $C$ is obtained when $f^c \leq \frac{\beta}{\alpha}f^{m}$, the time model can be simplified to $(t^0+\frac{\beta}{f^c})$, which eases the calculation of $f^{m}_o$. 

Based on the above analysis, the original three-variable problem is transformed into an one-variable optimization problem. Reducing the problem dimension is vital to speeding up the computation. Since the GPU voltage usually has a narrow range, we can conduct a grid search on it and derive the optimal core and memory frequencies in practice.

\subsection{Modeling the performance and power of GPU DVFS}
Theorem 1 allows the solution of Eq. \eqref{eq:argmin_e} via a single-variable optimization problem. The next step is estimating parameter values in the cost model for a GPU application. 
Unlike prior work \cite{amd_hpca2015,dvfs_aware_parco2019,ptx2019,taco2021_portable} directing predictions towards performance and power or the scaling ratio compared to the default DVFS setting, we suggest estimating model parameters using both DCGM metrics (dynamic information) and PTX instructions (static information) with a machine learning algorithm.

\subsubsection{Feature Processing}
We selected eight key metrics from the DCGM profiling tool closely tied to GPU kernel activities (Table \ref{tab:dcgm_features}), the values of which range between 0 and 1. As discussed in \cite{tpds2020_gpudvfs}, these metrics encompass the crucial factors essential for GPU performance modeling. 

Regarding PTX instructions, we consider three categories: instruction type, data type, and memory space (Table \ref{tab:ptx_features}). These include instruction types defined in the PTX ISA, basic instruction operand types, and all the types in the GPU memory hierarchy. We parse the PTX source code to count each instruction type and normalize each value by their category's total instructions.
\begin{table}[ht]
	\centering
	\caption{DCGM Features used in our GPU DVFS model} \label{tab:dcgm_features}
	\begin{tabular}{|c|p{2.5in}|} \hline
		Name & Descriptions \\ \hline \hline
		SMACT & The ratio of cycles an SM has at least one warp assigned \\ \hline
		SMOCC & The ratio of number of warps resident on an SM \\ \hline
		TENSO & The ratio of cycles the any tensor pipe is active \\ \hline
		DRAMA & The ratio of cycles the device memory interface is active sending or receiving data \\ \hline
		FP64A & Ratio of cycles the fp64 pipe is active \\ \hline
		FP32A & Ratio of cycles the fp32 pipe is active \\ \hline
		FP16A & Ratio of cycles the fp16 pipe is active \\ \hline
		INTAC & Ratio of cycles the integer pipe is active \\ \hline
	\end{tabular}
\end{table}

\begin{table}[ht]
	\centering
	\caption{PTX Features used in our GPU DVFS model} \label{tab:ptx_features}
	\begin{tabular}{|c|c|l|} \hline
		Type & Count & Examples \\ \hline \hline
		instruction & 101 & add, sub, fma, bra \\ \hline
		data & 17 & .u16, .f32, .f64 \\ \hline
		memory & 8 & .global, .shared, .local, .reg \\ \hline
	\end{tabular}
\end{table}

\subsubsection{Training Scheme}
Contrasting previous methods \cite{amd_hpca2015,icpp2019_predictable,ptx2019,tpds2020_gpudvfs} that predict absolute values or scaling factors, we propose to estimate the parameters in Eq. \eqref{eq:energy_model}, which can be used by Theorem 1 to derive the best configuration. This approach leverages efficient machine learning algorithms like Randomized Trees Regression \cite{taco2021_portable} and shallow neural networks \cite{ptx2019}, with an explicit formulation simplifying the learning process.
We utilize a multi-layer perceptron for estimating the energy model parameters, consisting of five layers including one input, three hidden layers (using a sigmoid activation function) with empirically set neurons (100, 50, 25), and one output layer. Hyper-parameters like batch size and learning rate are optimized with grid search and overfitting is prevented through three-fold cross-validation.
Practically, parameters in Eq. \eqref{eq:energy_model} are obtained by collecting data samples under all frequency settings for each GPU. Linear regression is then used to fit the power model and piecewise linear regression for the performance model, with the average regression absolute percentage error within 2\%. Our neural network is trained to estimate these model parameters, providing accurate predictions for realistic benchmarks.

\section{Experiments} \label{sec:Ex}
\subsection{Experimental Setup}

We validate the proposed DSO model on the Tesla V100 GPU. Experiments are performed on a Linux Ubuntu 18.04, with CUDA 11.5 and Nvidia Driver v515. Notice that the memory frequency of V100 is fixed since they cannot be tuned under our experimental environments. We change the GPU operating DVFS settings by \textsl{nvidia-smi} \cite{nvidia-smi} with the flag "-lgc" (for core frequency) and "-lmc" (for memory frequency). In our experiments, we tune no less than 10 frequency options, from 705 to 1380 MHz, to collect sufficient data samples to train the model.

To obtain all the data samples to train the machine learning algorithms for predicting the GPU DVFS model parameters, we execute each benchmark in \cite{gpumembench2016,mixbench2017} on the GPUs at all the available frequency configurations. By tuning the ratio of different instructions in each application, we obtain totally 138 GPU benchmarks of different operational intensity values.
After that, the accuracy of the estimated models is tested on 20 realistic benchmarks from CUDA SDK 11.5 and Rodinia. These benchmark applications cover a wide range of execution patterns, such as DRAM intensive, L2 cache intensive, shared memory intensive and computation intensive. Notice that the testing set is not used to train the models. Our model can perform well generalization on unseen GPU applications.

To build the DSO model, we obtain the necessary PTX instructions, DCGM metrics, and power samples for each benchmark. The "\textsl{-ptx}" flag of the nvcc compiler is used for PTX instructions, adjusting the PTX ISA version for different GPU architectures with "\textsl{-gencode=arch=compute\_XX,code=compute\_XX}" (in the case for the Tesla V100, which has a compute capability of 7.0, "XX" is 70). To gather DCGM metrics, a daemon thread runs "\textsl{dcgmi dmon -e \$\{METRIC\_LIST\}}" in loop during benchmark execution. A second daemon thread using NVML \cite{NVML} gathers power consumption data. Each benchmark's average power consumption is calculated from these samples. These efficient, non-intrusive methods demonstrate DSO's practical flexibility and efficiency.

\subsection{Model Accuracy}
\begin{figure}[ht]
	\centering
	\includegraphics[width=0.42\textwidth]{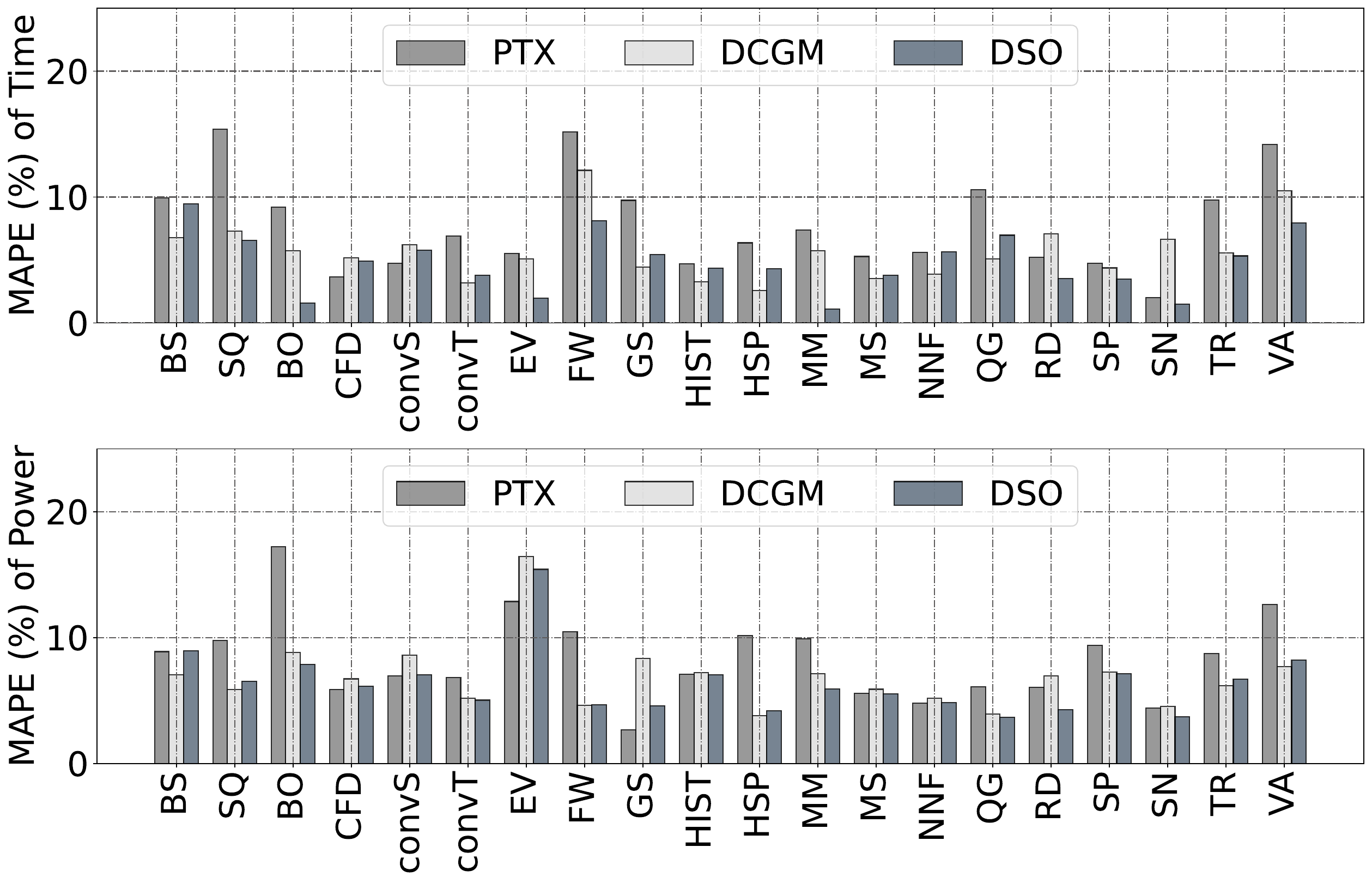}
	\caption{The prediction accuracy on Tesla V100.}\label{fig:v100_accuracy}
\end{figure}

Table \ref{tab:compare_exist} showcases our DSO in comparison to other studies, highlighting the modeling overhead and accuracy. Relying solely on static PTX code analysis proves challenging for execution time prediction, highlighting the importance of dynamic tools like nvprof and DCGM for accurate estimates. The DSO integrates DCGM for practical, high-quality, and low overhead estimation.
The performance and power estimation accuracy of DSO on the V100 GPU are manifested in Figure \ref{fig:v100_accuracy}. To emphasize the benefit of fusing dynamic and static information, DSO is compared with methods using either PTX features or DCGM metrics alone. DSO consistently yields low prediction errors for execution time (0.5\%-9.5\%) and power consumption (1.3\%-7\%). It outperforms the others with an average MAPE of 4.6\% (vs. PTX's 7.8\% and DCGM's 5.7\%) for performance and 4.9\% for power (vs. PTX's 8.3\% and DCGM's 6.9\%). It's observed that combining PTX and DCGM improves the accuracy drastically for several GPU applications, confirming the utility of runtime utilization information for capturing DVFS effects. This underlines the rationale behind DSO's design, using PTX features as a supplement to DCGM metrics for enhanced prediction.
\begin{table}[ht]
	\centering
	\caption{Comparison against the exisiting studies on Volta GPUs. The MAPE is measured for time (T) and power (P).} 
	\label{tab:compare_exist}
 		\addtolength{\tabcolsep}{-0.1pt}
		\resizebox{\linewidth}{!}{	\begin{tabular}{cccc} \hline
		Paper & Input & Modeling & Mean Average  \\ 
		& Source & Overhead &  Precision Error (MAPE) \\ \hline
		J. Guerreiro \cite{ptx2019} & PTX & low & 16.7(T)/6.7(P) \\ \hline
		Q. Wang \cite{tpds2020_gpudvfs}	 & nvprof & high & 8.8\%(T)/4.2\%(P) \\ \hline
		L. Braun \cite{taco2021_portable} & PTX & low & 11\%(T)/\textbf{2.3\%(P)}* \\ \hline
		DSO (ours)  & DCGM \& PTX & low & \textbf{4.6\%(T)}/4.9\%(P)  \\ \hline
	\end{tabular}
       }
       \scriptsize{*168 realistic kernels are used for cross-validation model training, which results in much better accuracy than others.}
\end{table}
\subsection{Energy Efficiency}
\begin{figure}[ht]
	\centering
	\includegraphics[width=0.42\textwidth]{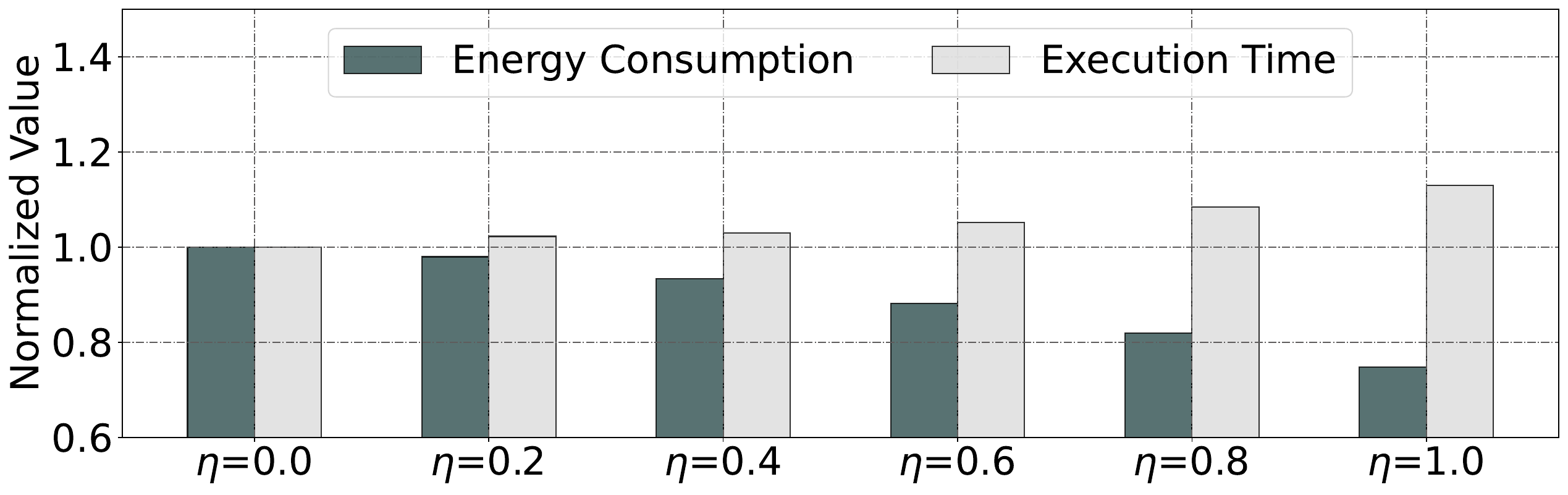}
	\caption{The energy and performance compared to the default setting on V100.}\label{fig:v100_ee_with_eta}
\end{figure}

Once predictive model parameters are garnered, the optimal DVFS configuration for minimal energy consumption is derived (Section \ref{subsec:solution}). Figure \ref{fig:v100_ee_with_eta} displays these results as average values over 20 real applications.
We examined the impact of varying $\eta$ values, which signal the preference level for energy efficiency. An $\eta$ of 1 indicates maximum preference for low energy consumption.
As $\eta$ increases, execution time expands, but energy consumption decreases. This results from additional opportunities for energy efficiency allowed by a higher $\eta$. Within performance loss limits (e.g., 5\%), our DSO offers a suitable $\eta$ selection according to these requirements.
For the V100 GPU architecture, we advise using $\eta$ = 0.8, conserving energy consumption by approximately 19\%.

\section{Conclusion} \label{sec:cc}
We introduce DSO, an innovative framework for modeling and optimizing GPU performance, power, and energy with DVFS. DSO uses a parameterized theoretical model considering DVFS effects to simultaneously optimize performance and energy efficiency. Combining static PTX and dynamic DCGM information, DSO accurately estimates model parameters, showing improved prediction accuracy over using either source independently. Tested on the Volta GPU, DSO models yield accurate results for unencountered real GPU applications. Furthermore, DSO facilitates balancing performance and energy efficiency. Leveraging DSO’s optimal configurations can enhance GPU energy efficiency by approximately 20\% with no more than a 5\% performance loss.

\section*{Acknowledgments}
This research was supported by the National Natural Science Foundation of China (No. 62302126), the Shenzhen Science and Technology Program (No. RCBS20221008093125065, No. JSGGKQTD20221101115655027) and the Peng Cheng Laboratory Project (Grant PCL2021A13). 

\bibliographystyle{IEEEtran}
\bibliography{arxiv}

\end{document}